# Interaction of internal solitary waves with long waves within the rotation modified Benjamin–Ono equation

R.H.J. Grimshaw[a], N.F. Smyth[b] and Y.A. Stepanyants[c,d*)]

[a] Department of Mathematics, University College London, Gower Street, London, WC1E 6BT, UK;
[b] School of Mathematics, University of Edinburgh, James Clerk Maxwell Building, The King's Buildings, Peter Guthrie Tait Road, Edinburgh, Scotland, EH9 3FD, UK;
[c] Faculty of Health, Engineering and Sciences, University of Southern Queensland, Toowoomba, QLD, 4350, Australia;
[d] Department of Applied Mathematics, Nizhny Novgorod State Technical University n.a. R.E. Alekseev, Nizhny Novgorod, Russia.

Abstract

Interaction of a solitary wave with a long background wave is studied within the framework of rotation modified Benjamin–Ono equation describing internal waves in a deep fluid. With the help of asymptotic method, we find stationary and nonstationary solutions for a Benjamin–Ono soliton trapped within a long sinusoidal wave. We show that the radiation losses experienced by a soliton and caused by the Earth rotation can be compensated by the energy pumping from the background wave. However, the back influence of the soliton on the background wave eventually leads to the destruction of the coherent structure and energy dispersion in a quasi-random wave field.

*) The corresponding author: Yury.Stepanyants@usq.edu.au

## 1. Introduction

In the paper by Ostrovsky & Stepanyants [2016] the interaction of solitary waves with the background long waves in the rotation modified Korteweg–de Vries (alias Ostrovsky equation) was studied. It was shown using asymptotic analysis and numerically that despite of "antisoliton theorem" [Leonov, 1981; Galkin & Stepanyants, 1991] that prohibits existence of stationary localised solitary waves, although they may exist on a variable background. In this paper we consider a similar problem for the rotation modified Benjamin–Ono (BO) equation [Grimshaw, 1985] describing, for example, internal waves on an oceanic pycnocline (sharp density interface). We study the interaction of a solitary wave (an algebraic BO soliton) with a long background wave. In the dimensional variables the basic equation reads:

$$\frac{\partial}{\partial x}\left(\frac{\partial \upsilon}{\partial t} + c_0 \frac{\partial \upsilon}{\partial x} + \alpha \upsilon \frac{\partial \upsilon}{\partial x} + \frac{\beta}{\pi}\frac{\partial^2}{\partial x^2}\wp\int_{-\infty}^{+\infty}\frac{\upsilon(\xi,t)}{\xi-x}d\xi\right) = \gamma\upsilon, \qquad (1)$$

where the coefficients $c_0$, $\alpha$, $\beta$, and $\gamma$ depend on the environmental parameters (background stratification, water depth, etc.). This equation can be considered as the generalisation of the classical Benjamin–Ono equation [Benjamin, 1967; Ono, 1976], which follows from Eq. (1) when $\gamma = 0$. In particular, in a density stratified fluid, with an upper layer $-d < z < 0$ of a variable density $\rho_0(z)$, and an infinitely deep lower layer of constant density $\rho_1$, the equation determining the modal function in the upper layer is [Grimshaw, 1985]:

$$c_0^2 \frac{d}{dz}\left(\rho_0(z)\frac{d\varphi}{dz}\right) + \rho_0(z)N^2(z)\varphi = 0, \qquad (2)$$

where $N^2(z) = -(g/\rho_0)(d\rho_0/dz)$. This equation should be augmented by boundary conditions:

$$c_0^2\, d\varphi/dz = g\varphi, \quad \text{at} \quad z=0; \quad \text{and} \quad d\varphi/dz = 0, \quad \text{at} \quad z=-d. \qquad (3)$$

Without loss of generality we can set $\varphi(z = -d) = 1$. Then the coefficients of Eq. (1) are:

$$\alpha = \frac{3}{I}\int_{-d}^{0}\rho_0(z)c_0^2\left(\frac{d\varphi}{dz}\right)^3 dz; \quad \beta = \frac{\rho_1 c_0^2}{I}\varphi^2(-d); \quad I = 2\int_{-d}^{0}\rho_0(z)c_0\left(\frac{d\varphi}{dz}\right)^2 dz; \quad I = \frac{f^2}{2c_0}, \qquad (4)$$

where $f = 2\Omega \cos \phi$, is the constant Coriolis parameter, $\Omega$ is the frequency of Earth rotation, and $\phi$ is the geographic latitude.

For perturbations of infinitesimal amplitudes with u ~ exp[$i\,(\omega\,t - k\,x)$] we obtain from Eq. (1) the dispersion relation:

$$\omega = c_0 k - \beta k |k| + \frac{\gamma}{k}; \quad c_p \equiv \frac{\omega}{k} = c_0 - \beta |k| + \frac{\gamma}{k^2}; \quad c_g \equiv \frac{d\omega}{dk} = c_0 - 2\beta |k| - \frac{\gamma}{k^2}, \tag{5}$$

where $c_p$ and $c_g$ are the phase and group velocities respectively.

Equation (1) is, apparently non-integrable, and even its stationary solutions are not known thus far in the analytic form. Particular numerical and approximate solutions describing adiabatic solitary wave decay under the influence of various dissipative mechanisms were constructed in Ref. [Grimshaw et al., 2018]. Because Eq. (1) contains two dispersive terms respectively proportional to $\beta$ and $\gamma$, one can expect that similarly to the Ostrovsky equation it can have a stationary solution representing a solitary wave riding on the crest or trough of a long background wave [Gilman et al., 1995; Chen & Boyd, 2001; Boyd & Chen, 2002]. In particular, if the solitary wave amplitude is not properly matched with the amplitude of the background wave, the solitary wave can travel along the long wave periodically accelerating and decelerating, growing and decaying (see, for example, [Gilman et al., 1996] where a similar situation occurs within the Ostrovsky equation). The solitary wave can be also trapped within the periodic background wave and oscillate around an equilibrium state near the trough of background wave [Ostrovsky & Stepanyants, 2016].

In the sequel we present an asymptotic theory describing the dynamics of a BO algebraic soliton on a long background wave. The background wave is taken as one of the particular solutions of the reduced rotation modified BO equation (1) with $\beta = 0$. As shown in [Ostrovsky, 1978; Ostrovsky & Stepanyants, 1989; Grimshaw et al., 1998; Boyd, 2005; Stepanyants, 2006], there exists a family of exact periodic solutions of such reduced equation. After giving general relationships in Sec. 2, we consider solitary wave interaction with a periodic background wave in Sec. 3. Then, in Sec. 4 we present the results of direct numerical simulation of solitary wave interaction with a periodic background wave and discuss the results obtained in Sec. 5.

**2. Interaction of a solitary wave with a long background wave.**

Solutions of the rotation modified BO equation (1) essentially depend on the relative sign of the dispersive coefficients $\beta$ and $\gamma$. Similarly to the Ostrovsky equation, it can be proven that for $\beta\gamma > 0$ (the "oceanic case"), localised solitary waves do not exist [Leonov, 1981; Galkin & Stepanyants, 1991]. In the opposite case $\beta\gamma < 0$ the "antisoliton theorem" is not valid, and solitary waves, apparently, can exist, but in this paper we will consider only the usual oceanic case with $\beta\gamma > 0$.

Equation (1) is, apparently nonintegrable, but it possesses at least two integrals of motion; one of them is the "zero mass" integral:

$$M \equiv \int \upsilon(x,t)dx = 0. \tag{6}$$

The integration here is taken either over the wave period for periodic waves or over the entire $x$-axis for localised solutions. When there is no rotation, $\gamma = 0$, the BO equation possesses an infinite number of integrals of motion [Ablowitz & Segur, 1981], and the "wave mass" integral $M$ can be an arbitrary constant which is determined by initial conditions. In our case, Eq. (6) is not just the integral of motion, but rather a constraint which demands that initial conditions must be consistent with the zero-mass condition.

To proceed further, it is convenient to reduce Eq. (1) to the dimensionless form by means of the transformations:

$$x' = \left(\frac{\gamma}{\beta}\right)^{1/3}(x - c_0 t), \quad t' = \beta\left(\frac{\gamma}{\beta}\right)^{2/3} t, \quad u = \frac{\alpha}{\beta}\left(\frac{\beta}{\gamma}\right)^{1/3}\upsilon. \tag{7}$$

Then, Eq. (1) can be presented as (the primes in new variables $x$ and $t$ are further omitted):

$$\frac{\partial}{\partial x}\left(\frac{\partial u}{\partial t}+u\frac{\partial u}{\partial x}+\frac{1}{\pi}\frac{\partial^2}{\partial x^2}\wp\int_{-\infty}^{+\infty}\frac{u(\xi,t)}{\xi-x}d\xi\right)=u. \tag{8}$$

Following Ref. [Ostrovsky & Stepanyants, 2016], we seek a solution to this equation in the form $u(t, x) = u_1(t, x) + u_2(t, x)$, where $u_1(t, x)$ is a smooth periodic background wave with the wavelength $\Lambda$, and $u_2(t, x)$ is a representation of a perturbed BO soliton with slowly varying amplitude and width (see below). If the characteristic soliton width $\Delta$ is much smaller than the wavelength $\Lambda$ of the background wave, the equations for these parts can be approximately separated. First we assume that function $u_1(t, x)$ is given that is, $u_1 = u_1(s = x - ct)$, where $c$ is a constant wave speed. For our present purposes we assume that this periodic wave is a long wave and so the linear dispersive term with the integral in Eq. (8) can be omitted. It then satisfies the reduced Ostrovsky equation (see, for example, [Ostrovsky, 1978; Stepanyants, 2006] and references therein):

$$\frac{d^2}{ds^2}\left(\frac{1}{2}u_1^2-cu_1\right)=u_1. \tag{9}$$

The shape of stationary solution $u_1(s)$ can vary from a small-amplitude sinusoidal wave to a limiting periodic wave in the form of a sequence of parabolic arcs as shown in Fig. 1 [Ostrovsky, 1978; Ostrovsky & Stepanyants, 1989; Grimshaw et al., 1998a; Boyd, 2005; Stepanyants, 2006]; all waves of this family have zero mean value.

As mentioned above, we intend that $u_2(t, x)$ in the trial solution describes a relatively narrow BO soliton with a characteristic width $\Delta \ll \Lambda$, which in the asymptotic limit as rotation is ignored, would satisfy Eq. (8) with a zero right-hand side. To derive the equation for the evolution off $u_2(t, x)$ under the influence of rotation, we first change the reference frame to $s = x - ct$ so that Eq. (8) becomes:

$$\frac{\partial u_2}{\partial t}+\left[u_1(s)-c\right]\frac{\partial u_2}{\partial s}+u_2\frac{\partial u_2}{\partial s}+\frac{1}{\pi}\frac{\partial^2}{\partial s^2}\wp\int_{-\infty}^{+\infty}\frac{u_2(\xi,t)}{\xi-s}d\xi=-\int_s^{\infty}u_2(\xi,t)d\xi-\frac{du_1}{ds}u_2. \tag{10}$$

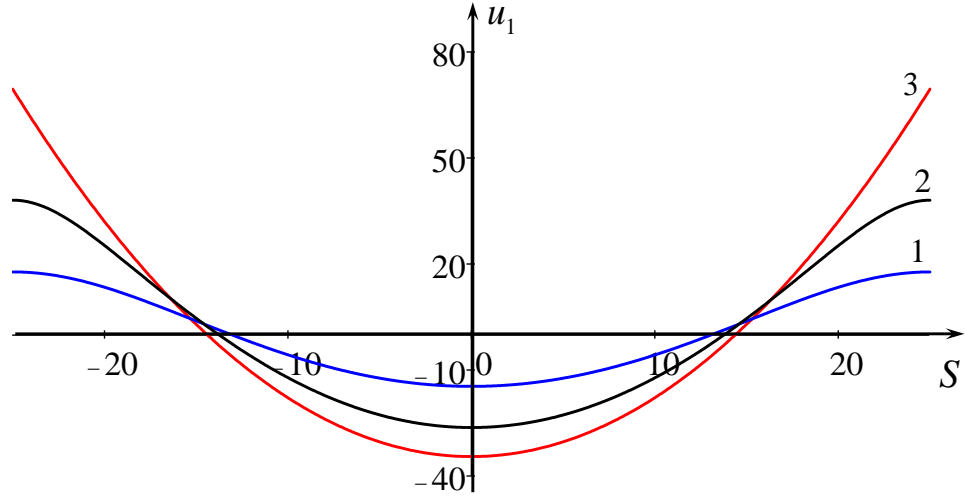

Fig. 1. (colour online). A family of zero-mass stationary periodic solutions of the reduced Ostrovsky equation (9) with $\Lambda = 50$. Line 1 pertains to quasi-sinusoidal wave with $c = 64$, line 2 illustrates a nonlinear wave with $c = 66$, and line 3 represents a periodic sequence of parabolic arcs with $c = (25/3)^2 \approx 69.4$.

We further assume that the solitary wave is localised at a location $s = S(t)$ and rewrite Eq. (10) as:

$$\frac{\partial u_2}{\partial t} + \left[u_1(S) - c\right]\frac{\partial u_2}{\partial s} + u_2 \frac{\partial u_2}{\partial s} + \frac{1}{\pi}\frac{\partial^2}{\partial s^2} \wp \int_{-\infty}^{+\infty} \frac{u_2(\xi,t)}{\xi - s} d\xi$$

$$= -\int_s^{\infty} u_2(\xi,t) d\xi - \frac{\partial}{\partial s}\left\{\left[u_1(s) - u_1(S)\right]u_2\right\}. \tag{11}$$

For a relatively short solitary wave the right-hand side of this equation can be treated a small perturbation term, and when this term is neglected, Eq. (11) has an exact soliton solution,

$$u_2 = \frac{A}{1 + \left[(s-S)/\Delta\right]^2}, \tag{12}$$

where $\Delta = 4/A$ and

$$\frac{dS}{dt} = u_1(S) - c + \frac{A}{4}. \tag{13}$$

When this solitary wave travels along the background long wave its amplitude $A$, slowly vary with time. To derive the equation for the amplitude variation, one can use one of the known regular asymptotic methods, described, for instance, in [Grimshaw et al., 1998b; Ostrovsky & Gorshkov, 2000; Grimshaw & Helfrich, 2008; Obregon & Stepanyants; 2012]. Here we use an alternative simplified approach similar to that developed in [Gorshkov & Ostrovsky, 1981] for the description of KdV solitary wave interaction with a sinusoidal pump wave in the electro-magnetic transmission line and based on the phase and energy balance.

The equation for the solitary wave amplitude follows from the energy balance equation [Ostrovsky & Gorshkov, 2000; Grimshaw & Helfrich, 2008]. Multiplying Eq. (11) by $u_2$ and integrating over $s$ within the interval $[-L/2, L/2]$ in the vicinity of soliton maximum, where $\Delta \ll L \ll \Lambda$, we obtain (cf. [Ostrovsky & Stepanyants, 2016]):

$$\frac{d}{dt}\int_{-L/2}^{L/2} u_2^2 ds \approx -\left(\int_{-L/2}^{L/2} u\, ds\right)^2 - 2\frac{du_1}{ds}\int_{-L/2}^{L/2} u_2^2 ds. \qquad (14)$$

Here it was taken into account that the gradient of the background wave $du_1/ds$ is approximately constant in the neighbourhood of the solitary wave, whose field rapidly decays away from its centre, so that $u_2(\pm L/2) \approx 0$ (i.e., one can set $L = \infty$). Substituting the soliton solution (12) for $u_2(t, x)$, after integration we obtain:

$$\frac{dA}{dt} = -8\pi - A\frac{du_1(s)}{ds}\bigg|_{s=S(t)}. \qquad (15)$$

The set of equations (13) and (15) defines the variation of the amplitude $A$ and phase $S$ in time. To get some insight of the possible effects, we consider below only the limiting cases of the background small-amplitude sinusoidal wave.

## 3. Interaction of a BO solitary wave with a sinusoidal wave

Let us assume that the long background wave is sinusoidal:

$$u_1(s) = U_0 \sin\left(ks - \frac{\pi}{2}\right) \qquad (16)$$

with the period $\Lambda = 2\pi/k$, amplitude $U_0 \ll U_{lim} \equiv \Lambda^2/36$, and phase speed $c = 1/k^2$. The restriction on the wave amplitude follows from the comparison with the amplitude of a limiting wave representing a sequence of parabolic arcs shown in Fig. 1, and the phase speed of the sinusoidal wave follows from the linearized Eq. (8) when both the nonlinear term and second-order derivative are neglected.

**3.1 The conservative case**

Consider first the solitary wave dynamics in the variable background field, ignoring the effect of radiative losses, which the solitary wave experiences due to the influence of large-scale dispersion. The effect of such radiation has been studied by [Grimshaw et al., 2018], who showed that in finite time a BO soliton experiences terminal decay, when it transforms to a radiating wave train (a similar process has been studied for the KdV soliton within the framework of the Ostrovsky equation [Grimshaw et al., 1998a]).

Here we study the influence of the background wave only on the dynamics of a solitary wave neglecting for the moment the radiative losses. Substituting the solution (16) for the background wave $u_1(S)$ into Eqs. (13) and (15), omitting the radiative loss term $-8\pi$ in (15), we obtain a set of two equations for $S$ and $A$:

$$\frac{dS}{dt} = \frac{A}{4} - U_0 \cos kS - \frac{1}{k^2}, \tag{17}$$

$$\frac{dA}{dt} = -AU_0 k \sin kS. \tag{18}$$

This system is conservative with the Hamiltonian $H(A, S)$

$$H(A,S) = \frac{A^2}{8} - AU_0 \cos kS - \frac{A}{k^2}, \tag{19}$$

which is a conserved quantity. Here $S = 0$ corresponds to one of the troughs of the background wave playing a role of energy pump for the solitary wave. Consider first the stationary points of this dynamical system. In general, there are four such points within each period of sinusoidal wave:

$$A_1 = 4\left(\frac{\Lambda^2}{4\pi^2} + U_0\right), \quad S_1 = 0 \pm n\Lambda; \tag{20}$$

$$A_2 = 4\left(\frac{\Lambda^2}{4\pi^2} - U_0\right), \quad S_2 = -\Lambda/2 \pm n\Lambda. \tag{21}$$

$$A_{3,4} = 0, \ kS_{3,4} = \cos^{-1}(-1/k^2 U_0). \tag{22}$$

Here $n$ is an integer. Two of these stationary points, $A_{1,2}$, correspond to solitary waves of non-zero amplitude. According to (21), the amplitude $A_2$ exists only when $\Lambda^2/U_0 > (2\pi)^{2}$[2]. Solutions with zero amplitudes cannot be considered here because for such solitons the characteristic widths become infinite; this violates our assumption $\Delta \ll \Lambda$ or $A \gg 4/\Lambda$. Figure 2 illustrates stationary solitary waves riding on the sinusoidal background wave.

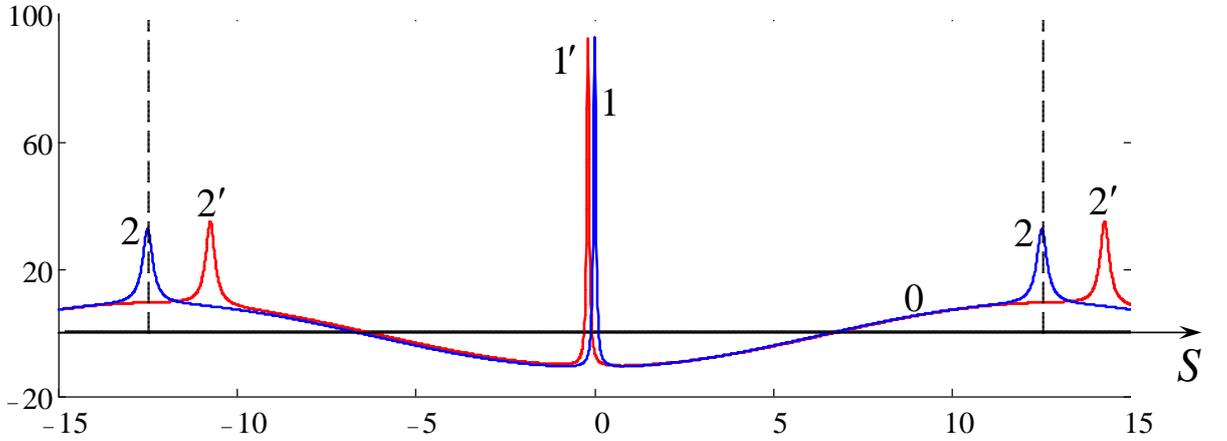

Fig. 2 (colour online). One spatial period of the background sinusoidal wave (12) with $\Lambda = 25$ and amplitude $U_0 = 10$ (line 0) with the BO solitons sitting on the troughs (1) and on the crests (2). Lines 1′ and 2′ pertain to stationary solitons with the radiative losses (see below). Vertical dashed lines show the period of the background wave.

The phase portraits are then given by $H(A, S)$ = constant, which can be evaluated at the point $A = A_0$, $S = S_0$, and are shown in Fig. 3. There is a family of closed trajectories describing solitary wave oscillations in the potential well of the background wave around its minima. There is also a family of open trajectories describing solitary waves travelling along the background wave either in the same direction with the background wave (the trajectories above line 1) or in the opposite direction (the trajectories below line 2; they are not shown). The separatrix consisting of two branches (lines 1 and 2) describes solitary wave which travels from one crest of the background wave to another one. One of the stationary points (20) of the centre-type is located at $S = 0$, i.e. in the trough of the background wave; the stationary points (21) of the saddle-type shown in the corners of the separatrices 1 and 2, at $S = \Lambda/2$, correspond to unstable equilibrium positions on the wave crests.

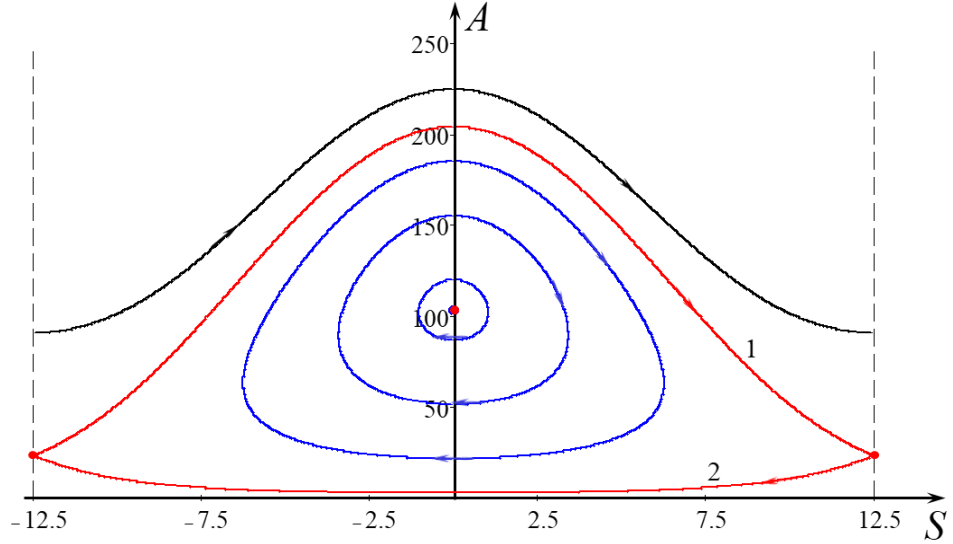

Fig. 3 (colour online). The phase portrait of the dynamical system (17), (18) in the conservative case with $\Lambda = 25$, $U_0 = 10$. Lines 1 and 2 are the separatrices. Vertical dashed lines show the boundary of one cell of the phase portrait which is periodic in the horizontal direction.

The period of oscillations around the central point can be easily estimated for relatively small amplitudes of oscillations. It follows from the dynamical system (17), (18) linearized around the equilibrium state (19) that the period of oscillations is:

$$T = \frac{2\pi \Lambda}{\sqrt{U_0 \left( \Lambda^2 + 4\pi^2 U_0 \right)}}. \tag{23}$$

Substituting here $\Lambda = 25$ and $U_0 = 10$ as in Fig. 3, we obtain $T \approx 1.56$.

**3.2 Influence of radiative losses on soliton dynamics**

Now let us take into consideration the radiative losses which the BO soliton experiences due to the rotational effect [Grimshaw et al., 2018]. In the case of sinusoidal background wave Eq. (18) is replaced by:

$$\frac{dA}{dt} = -AU_0 k \sin kS - 8\pi. \tag{24}$$

Although the set of equations (17) and (24) is not energy conserving, there is a Hamiltonian-like conserved quantity $H_{rl}(A, S)$ given by:

$$H_{rl}(A,S) = \frac{A^2}{8} - AU_0 \cos kS - \frac{A}{k^2} - 8\pi S . \qquad (25)$$

As before, the phase portrait is obtained from $H_{rl}(A, S)$ = constant and is shown in Fig. 4.

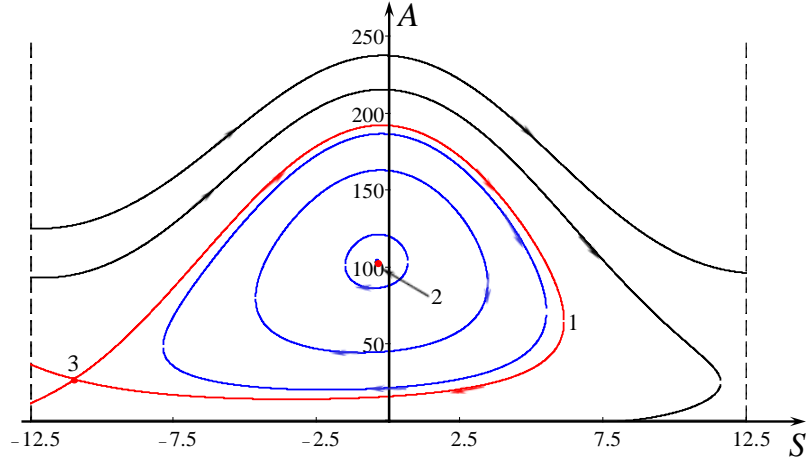

Fig. 4 (colour online). The phase portrait of the dynamical system (17), (24). Line 1 is the separatrix separating closed and unclosed trajectories. Points 2 and 3 represent equilibrium states. Vertical dashed lines show the boundaries of the main cell.

In contrast to the Ostrovsky equation [Ostrovsky & Stepanyants, 2016], the dynamical system (17), (24) is quasi-conservative. All its solutions corresponding to trajectories outside of a separatrix 1 in Fig. 4, eventually decay and vanish, although before vanishing the soliton amplitude can oscillate with $S$ and travel from one cell of phase plane to another. This corresponds to a BO solitary wave travelling on a sinusoidal background wave until it arrives in a certain period where it quickly decays. In the meantime, all trajectories within the separatrix are closed and represent the periodic oscillations around equilibrium point 2 in Fig. 4 as in the conservative case.

The equilibrium points of the dynamical system (17), (24) can be found by setting the time derivatives to zero. Then after some manipulation we obtain the equation to determine $S$:

$$\tan^4 \frac{kS}{2} + \frac{kU_0^2}{\pi}\left(\frac{1}{k^2 U_0} - 1\right)\tan^3 \frac{kS}{2} + 2\tan^2 \frac{kS}{2} + \frac{kU_0^2}{\pi}\left(1 + \frac{1}{k^2 U_0}\right)\tan \frac{kS}{2} + 1 = 0. \tag{26}$$

It can be shown that this quartic polynomial with respect to tan (kS/2) has only two real roots, which can be found analytically, in principle, but it is easier to find them numerically. Having these roots $S_{eq}$, one can find from Eq. (17) with zero left-hand side, the soliton amplitudes:

$$A_{eq} = 4\left(U_0 \cos kS_{eq} + \frac{1}{k^2}\right). \tag{27}$$

Two equilibrium points are shown in Fig. 4 for $U_0 = 10$ and $\Lambda = 2\pi/k = 25$; they are: $A_1 = 103.13$, $S_1 = -0.39$ (see point 2 in Fig. 4) and $A_2 = 26.32$, $S_2 = -10.95$ (see point 3 in Fig. 4). The former one is the centre, whereas the latter is the saddle. There is one trajectory in each cell which originates in the saddle point 3 makes a loop around the centre and returns back to the saddle in infinite time. Such trajectory corresponds to the unstable solitary wave solution.

The quartic polynomial (26) has real roots only if $U_0 \geq U_{cr} = 4\pi^2/\Lambda$, therefore two equilibrium positions can exist when this condition holds. When $U_0$ approaches the critical value from the top, the solitary wave amplitudes become equal and their equilibrium positions coincide at $S_{eq} = -\Lambda/4$, where the gradient of the background wave is maximal. In such position the energy pumping to the solitary waves from the sinusoidal background wave is maximal and compensates the radiative energy losses.

Figure 2 shows stationary solitary waves riding on the background sinusoidal wave of the same amplitude $U_0 = 10$, but with the radiative losses taken into account. Soliton 1′ is in the stable position, and soliton 2′ is in the unstable (saddle-type) position. As compared with the conservative case, the unstable solitary wave shifts forward from the crest and thus its phase becomes slightly greater than $\Lambda/2$, whereas the stable solitary wave shifts backward from the trough and its phase becomes slightly negative. This is caused by the energy exchange with the background wave, because only on the negative (frontal) slope of the sinusoidal wave a solitary wave can acquire energy from the background wave to compensate its energy loss due to radiation.

In Fig. 5 we show the near-critical situation when two solitary waves with the radiative effect taken into account are riding on a sinusoidal wave of a small amplitude $U_0 = 1.58$. In such a situation when the solitary waves are very close to each other, the interaction between them should be taken into account; this leads to further complication of description of soliton dynamics (more details can be found in Ref. [Ostrovsky & Stepanyants, 2019] for two KdV solitary waves interacting with the periodic background wave within the Ostrovsky equation.

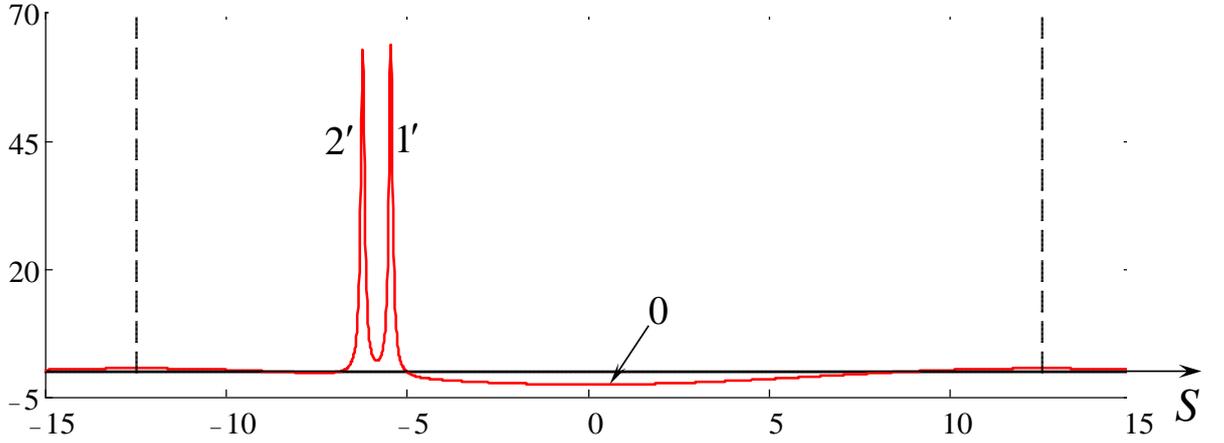

Fig. 5 (colour online). One spatial period of the background sinusoidal wave (16) with $\Lambda = 25$ and amplitude $U_0 = 1.28$ (line 0) with the BO solitary waves sitting at the equilibrium positions when the radiative losses are taken into account. Vertical dashed lines show the period of the background wave.

In Fig. 6 we illustrate the dependence of the stationary solitary wave amplitudes on the amplitude of the background sinusoidal wave for $\Lambda = 25$. Dashed lines 1 and 2 refer to the conservative case as in Eqs. (20) and (21) for solitary waves propagating on the troughs and crests, respectively. Lines 3 and 4 correspond to the case when the radiative losses are taken into account. In this case stationary solitary waves can exist only when the amplitude of the background wave is large enough, $U_0 \geq U_{cr} = 4\pi^2/\Lambda$. At the critical condition solitary waves with equal amplitudes formally are located at at $S_{eq} = -\Lambda/4$, where the background wave gradient is maximal (see Fig. 7).

When $U_0$ increases above the value $U_{cr}$, lines 3 and 4 rapidly approach the asymptotic lines 1 and 2 which represent solitary wave amplitudes $A_{1,2}$ in the conservative case as in Eqs. (20) and (21).

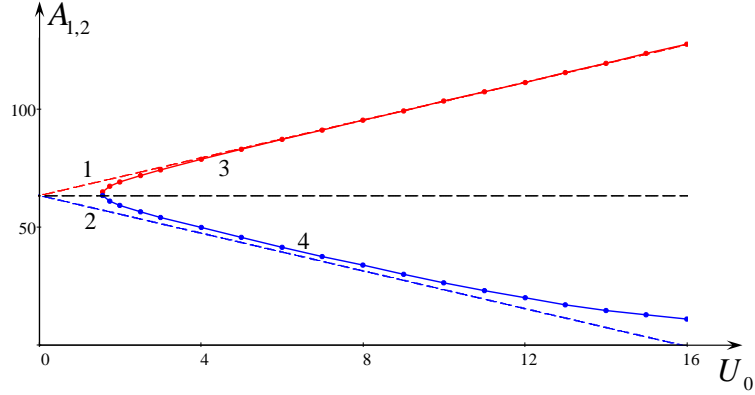

Fig. 6 (colour online). Stationary solitary wave amplitudes versus the amplitude of the background sinusoidal wave for $\Lambda = 25$ in the conservative case (lines 1 and 2) and when the radiative losses are taken into account (lines 3 and 4). Lines 1 and 3 pertain to solitary waves sitting in the troughs, and lines 2 and 4 – to the solitary waves sitting on the crest.

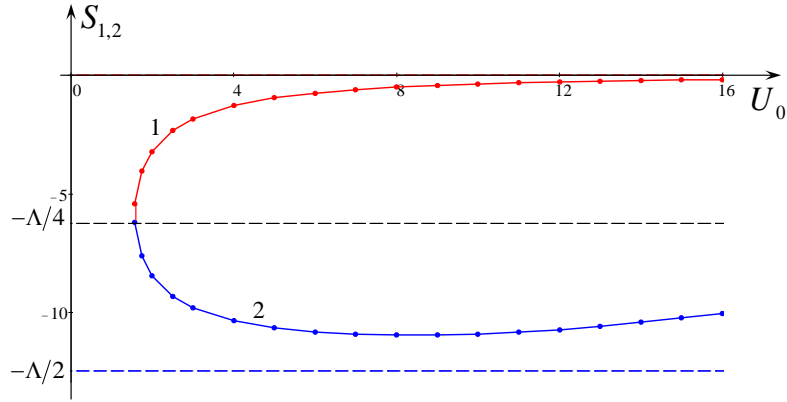

Fig. 7 (colour online). Positions of stationary solitons on the background sinusoidal wave against the amplitude $U_0$ for $\Lambda = 25$. Line 1 pertains to the soliton sitting next to the trough, and line 2 and – to the soliton sitting next to the crest. In the conservative case they would be sitting at $S = 0$ and $S = \Lambda/2$ respectively.

The stationary positions of solitary waves on the background wave which can be found from the roots of the quartic polynomial (26) are shown in Fig. 5 in terms of the dependences $S_{1,2}$ against $U_0$ for $\Lambda = 25$. The plots 6 and 7 are valid for amplitudes $U_0 \ll U_{lim} \equiv \Lambda^2/36$ (see after Eq. (16)); in the considered case of $\Lambda = 25$, $U_{lim} = (25/6)^2 \approx 17.36$.

The theory developed here describes the basic effects of the interaction between a solitary wave and a long sinusoidal wave. The main consequence is that, unlike the case of a solitary wave nonexistence in a homogeneous background, it can exist and be stable on a long wave, since the pumping of energy from the background wave can compensate the radiative losses. At the same time there are various additional factors which were ignored so far. Below we consider one of these factors, the inverse impact of a solitary wave on the background sinusoidal wave.

**3.3 Inverse impact of a solitary wave on the background sinusoidal wave**

In the above analysis the background wave was assumed given. However, in reality this wave undergoes attenuation due to energy consumption by the solitary wave. The energy of the background wave changes at the same rate as it is consumed by the solitary wave riding on it. We assume that there is only one solitary wave riding in each period of the background wave, and neglect the direct interaction between the solitary waves. In this case the energy balance equation analogous to (14) is, written for each period of the background wave:

$$\frac{d}{dt}\int_{-\Lambda/2}^{\Lambda/2} u_1^2 ds = -\int_{-L/2}^{L/2} u_2^2 \frac{du_1}{ds} ds \approx -\frac{du_1}{ds}\int_{-L/2}^{L/2} u_2^2 ds. \tag{28}$$

In the right-hand side of this equation the derivative of $u_1(s)$ is taken at the location of a solitary wave $s = S(t)$; it can be considered a constant here, because the solitary wave width is very small, and its shape resembles a Dirac delta-function. Substituting here the expressions (12) and (16) for $u_2$ and $u_1$, respectively, and integrating, we obtain the equation for the slowly varying amplitude of the background wave $U(t)$:

$$\frac{dU}{dt} = Ak^2 \sin kS . \tag{29}$$

Now the set of three equations, (17), (24), and (29), should be analysed, where in (17) and (24) we replace $U_0$ with $U(t)$. For convenience we collect all three equations together:

$$\frac{dS}{dt} = \frac{A}{4} - U\cos kS - \frac{1}{k^2}, \tag{17a}$$

$$\frac{dA}{dt} = -AUk \sin kS - 8\pi. \tag{24a}$$

$$\frac{dU}{dt} = Ak^2 \sin kS. \tag{29a}$$

This dynamical system of "1.5 degrees of freedom" has the first integral:

$$2\pi A(t) + \frac{1}{2} U^2(t) \Lambda = 2\pi A_0 + \frac{1}{2} U_0^2 \Lambda - 16\pi^2 t, \tag{30}$$

where the left-hans side represents the total energe of solitary wave (the first term) and one period of background sinusoidal wave (the second term). Equation (30) shows that the total energy of the wave field linearly decreases with time due to radiation losses experienced by the solitary wave. The solitary wave borows the energy from the backgound wave and radiates it further. If there is no the background wave, we obtain form Eq. (30) the decay law of soliton amplitude previously derived in [Grimshaw et al., 2018].

The dynamical system (17a), (24), and (29) is non-integrable, but can be easily investigated numerically. As follows from Eq. (30), the solitary wave amplitude in this case inevitably vanishes in a finite time which depends on the initial conditions, and there is no an equilibrium state in the system. A sample of numerical solution of the dynamical system (17a), (24), and (29) is shown in Fig. 8 for the following initial condition: $A_0 = 1140$, $S_0 = 0$, and $U_0 = 15$, $\Lambda = 100$.

As one can see, the solitary wave experiences a few oscillations around the trough of the background wave, then quickly decays with the linear rate in time in avarege (see frame a). However, the solutions becomes meaningless from the physical point of view when solitary wave amplitude approaches zero and its width increases. The amplitude of the background wave decays as expected, changes its phase by $\pi$, and eventually stabilises around the terminal value $|U_t| \approx 3.46$ (see horizonatl dashed line in frame c).

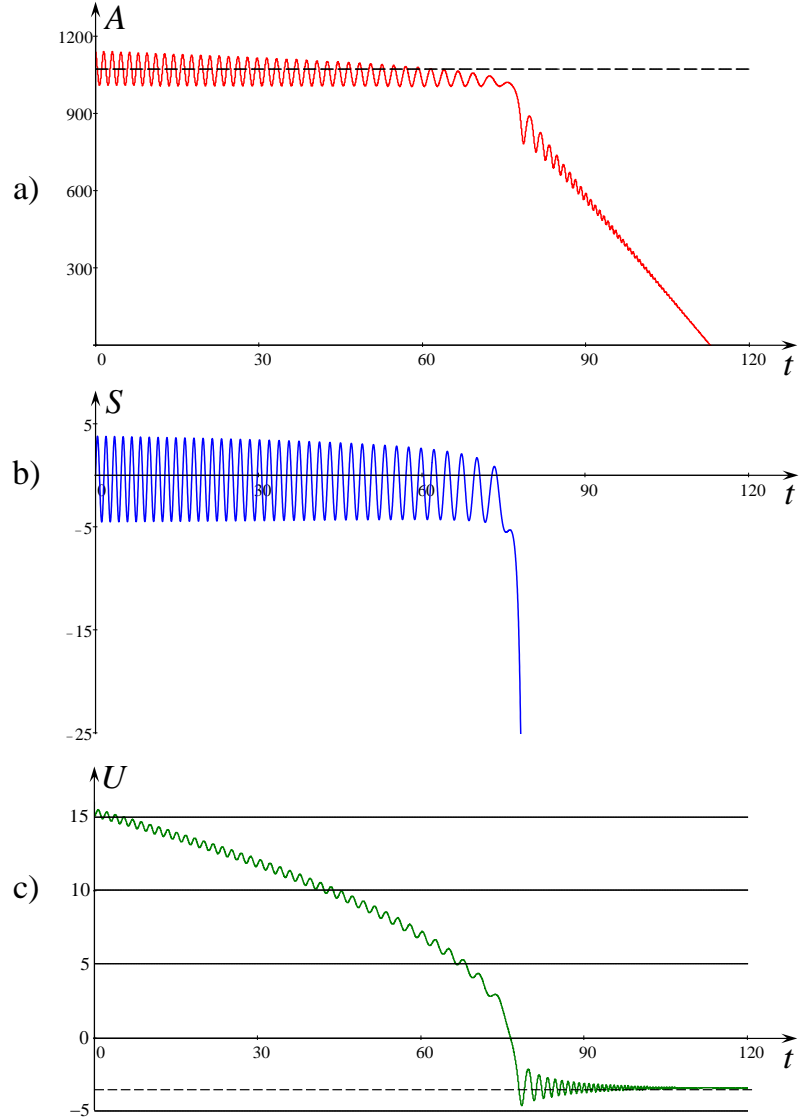

Fig. 8 (colour online). The dependences of the solitary wave amplitude $A(t)$ (frame a) and position $S(t)$ (frame b) on time. Dashed horizontal line in frame a) shows stationary solitary wave amplitude $A_1$ when the influence of solitary wave on the background wave is ignored. Frame c) shows the dependence of amplitude of the background sinusoidal wave on time. Dashed horizontal lines in this frame shows stationary amplitude of the sinusoidal wave when the solitary wave completely vanishes (the negative value implies that phase of sine wave was changed by $\pi$).

Figure 9 illustrates the typical phase trajectory in the three-dimensional phase space of the dynamical system (17a), (24a), and (29a). At the initial stage the solitary wave amplitude oscillates and decays within one period of the background wave and consumes energy to compensate the radiation losses. Then its amplitude drastically decreases. The process of solitary wave decay can also be demonstrated on the phase plane which is a projection of 3D phase space of Fig. 9 onto *SA*-plane as shown in Fig. 10 by line 2. In the same figure we show also the phase trajectory of the dynamical system (17a), (24a) when the amplitude of the background sinusoidal wave is assumemed constant, and the influence of solitary wave on the background wave is ignored (see line 1). In both cases for lines 1 and 2 the initial conditions were the same.

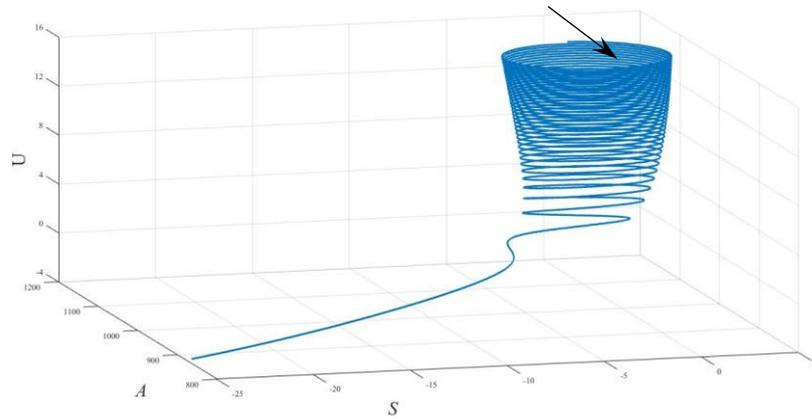

Fig. 9 (colour online). An example of the typical phase trajectory of the system (17a), (24a), and (29a) in 3D phase space for the particular initial condition: $U_0 = 15$, $A_0 = 1140$, $S_0 = 0$, and $\Lambda = 100$. Black arrow shows the starting point.

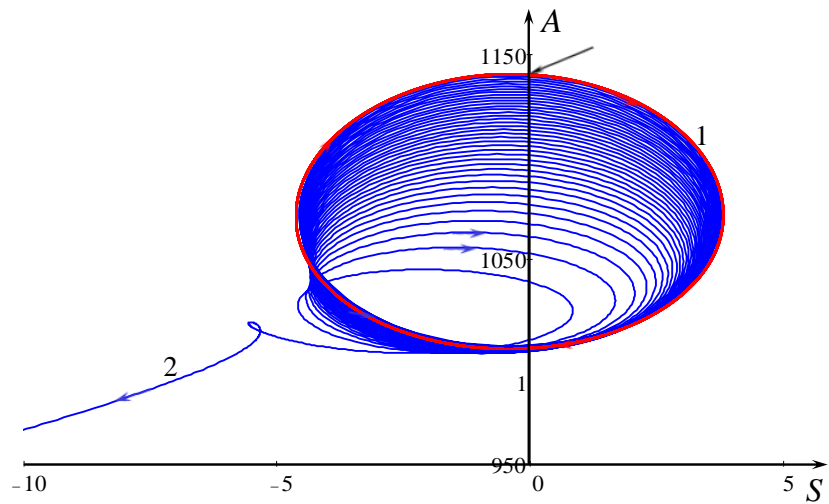

Fig. 10 (colour online). A projection of the typical phase trajectory shown in Fig. 9 onto the *SA*-plane. Black arrow shows the starting point. Line 1 represents the phase trajectory when the amplitude of background wave is fixed and line 2 is the phase trajectory when the amplitude of background waves varies with time.

## 4. Conclusion

In this paper, we have studied theoretically the process of interaction of internal BO soliton with a long background wave in a rotating ocean. By means of an asymptotic method, we have derived a set of approximate equations describing the amplitude and position of a soliton trapped in the background wave. The analysis of the set shows that there are equilibrium positions for a soliton when its energy loses due to radiation caused by the rotation effect are compensated by the energy pumping from the background wave. The soliton can experience decaying oscillations around the equilibrium position. However, the back influence of the soliton on the background wave eventually leads to the destruction of the coherent structure and energy dispersion in a quasi-random wave field. The theoretical results obtained here will be compared with the numerical data in the nearest future.